\documentclass[sigconf]{acmart}
\AtBeginDocument{%
  }

\copyrightyear{2025}
\acmYear{2025}
\setcopyright{acmlicensed}\acmConference[SIGIR '25]{Proceedings of the 48th
International ACM SIGIR Conference on Research and Development in
Information Retrieval}{July 13--18, 2025}{Padua, Italy}
\acmBooktitle{Proceedings of the 48th International ACM SIGIR Conference on
Research and Development in Information Retrieval (SIGIR '25), July 13--18,
2025, Padua, Italy}
\acmDOI{10.1145/3726302.3729933}
\acmISBN{979-8-4007-1592-1/2025/07}

\usepackage{enumitem}
\usepackage{multirow}
\usepackage{multicol}
\usepackage{subcaption}

\begin{document}

\title{Comprehensive List Generation for Multi-Generator Reranking}


\author{Hailan Yang}
\affiliation{%
  \institution{Kuaishou Technology}
  \city{Beijing}
  \country{China}}
\email{yanghailan@kuaishou.com}

\author{Zhenyu Qi}
\affiliation{%
  \institution{Kuaishou Technology}
  \city{Beijing}
  \country{China}}
\email{qizhenyu@kuaishou.com}

\author{Shuchang Liu}
\authornote{Corresponding author}
\affiliation{%
  \institution{Kuaishou Technology}
  \city{Beijing}
  \country{China}}
\email{liushuchang@kuaishou.com}

\author{Xiaoyu Yang}
\affiliation{%
  \institution{Kuaishou Technology}
  \city{Beijing}
  \country{China}}
\email{yangxiaoyu@kuaishou.com}

\author{Xiaobei Wang}
\affiliation{%
  \institution{Kuaishou Technology}
  \city{Beijing}
  \country{China}}
\email{wangxiaobei03@kuaishou.com}

\author{Xiang Li}
\affiliation{%
  \institution{Kuaishou Technology}
  \city{Beijing}
  \country{China}}
\email{lixiang44@kuaishou.com}

\author{Lantao Hu}
\affiliation{%
  \institution{Kuaishou Technology}
  \city{Beijing}
  \country{China}}
\email{hulantao@kuaishou.com}

\author{Han Li}
\affiliation{%
  \institution{Kuaishou Technology}
  \city{Beijing}
  \country{China}}
\email{lihan08@kuaishou.com}

\author{Kun Gai}
\affiliation{%
  \institution{Unaffiliated}
  \city{Beijing}
  \country{China}}
\email{gai.kun@qq.com}

\renewcommand{\shortauthors}{Hailan et al.}

\begin{abstract}
Reranking models solve the final recommendation lists that best fulfill users' demands.
While existing solutions focus on finding parametric models that approximate optimal policies, recent approaches find that it is better to generate multiple lists to compete for a ``pass'' ticket from an evaluator, where the evaluator serves as the supervisor who accurately estimates the performance of the candidate lists.
In this work, we show that we can achieve a more efficient and effective list proposal with a multi-generator framework and provide empirical evidence on two public datasets and online A/B tests.
More importantly, we verify that the effectiveness of a generator is closely related to how much it complements the views of other generators with sufficiently different rerankings, which derives the metric of list comprehensiveness.
With this intuition, we design an automatic complementary generator-finding framework that learns a policy that simultaneously aligns the users' preferences and maximizes the list comprehensiveness metric.
The experimental results indicate that the proposed framework can further improve the multi-generator reranking performance.
\end{abstract}


\maketitle

\section{Introduction}\label{sec: introduction}

\begin{figure}[t]
    \centering
    \includegraphics[width=\linewidth]{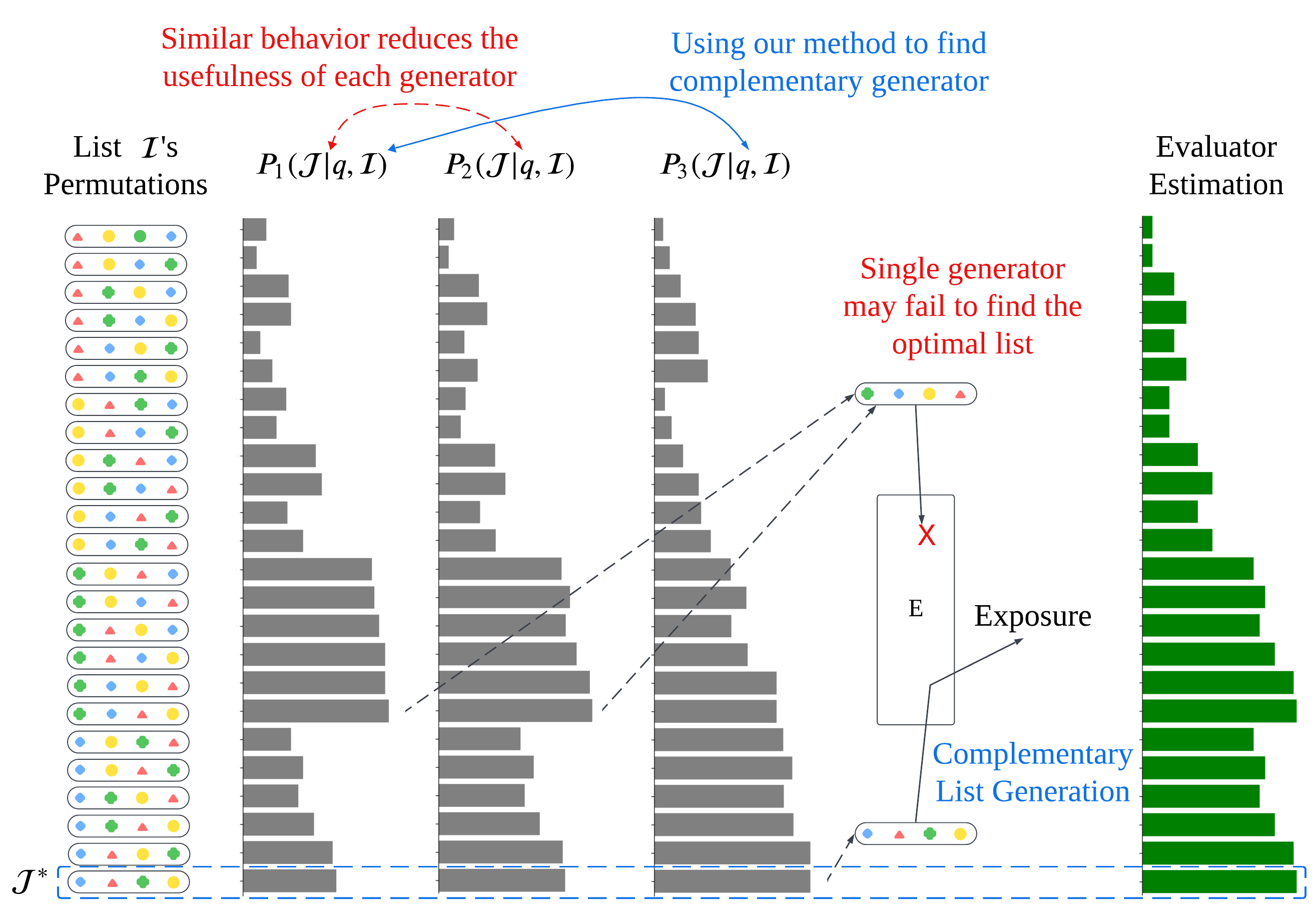}
    \caption{Intuitive example for the benefit of multi-generator design in reranking task and the motivation of list comprehensiveness. $P_1$, $P_2$, and $P_3$ are different generators. $E$ represents the efficient and accurate list reward estimator that select the best list as exposure. $\mathcal{I}$ and $\mathcal{J}^\ast$ corresponds to the initial list and the optimal list. $q$ denotes the user request.}
    \label{fig: intuition}
\end{figure}

Recommender systems filter information of interest to users for a wide range of web services such as e-commerce, news, social media, and video sharing platforms.
Practical recommendation settings usually involve large and dynamic candidate sets, which bring forth the challenge of accuracy-latency trade-off: sophisticated solutions cannot meet the latency demand due to the large candidate set, and simple models cannot meet the accuracy demand that directly influences user satisfaction.
As a countermeasure, the multi-stage design~\cite{wang2011cascade,qin2022rankflow,zheng2024full} has been proposed to iteratively refine and filter relevant items/contents that best align with the user's interests.
In the final reranking stage~\cite{liu2022neurerank} that determines the final exposure, the list provided from previous ranking stages is reordered before presentation to the user, considering the cross-item influence.
For example, an item with repeating information to a previous item may not satisfy the user's need for content diversity, but an item may become more attractive if the user has previously interacted with a complementary item.

Formally, the reranking task aims to find the optimal list that maximizes the user response taking into account the cross-item mutual influence.
Early studies~\cite{carbonell1998mmr,zhuang2018mutual} found that the mutual influence is closely related to the recommendation diversity and have proposed several heuristic-based techniques to balance the diversity-accuracy trade-off.
Later approaches~\cite{ai2018listwise,pei2019prm,gong2022edgererank}, denoted as the generator-only paradigm, adopted machine learning methods that learn a policy model that directly models the interactions among the candidates, learns how different rankings affect the list-level reward, and generates the ranking according to the optimum reward.
However, finding the best ranking is combinatorially complex, and the problem space grows exponentially with the size of the candidate set and the size of the output recommendation list.
Furthermore, it is unknown how the user would respond to an ordering different from the observed data sample, which drives the efforts to find extra guidance as reinforcement.
To mitigate these issues, several recent works~\cite{shi2023pier,ren2024nar} have proposed the two-step generator-evaluator (G-E) framework, where the generator selects a number of high-value candidate lists among all possible permutations, avoiding the exhaustive search in the list space; then the evaluator accurately estimates the utility/reward of each list, serves as a simulator that mimics the user's response for unseen lists, and picks the list with the maximum reward as the recommendation.
Yet, the naive list permutation process in the generator step of the G-E framework injects random noises, which do not guarantee accuracy during list exploration.
Furthermore, the single-generator design may still be ineffective even with the similarity-based exploration~\cite{shi2023pier} when the optimal list is outside of the local optima (e.g., The optimal region of $P_1$ in Figure \ref{fig: intuition} only covers the lists around the sub-optimal list in the ground truth provided by the evaluator).

In this work, we investigate an alternative solution to the reranking task, which overcomes the aforementioned limitations with a \textbf{multi-generator-evaluator (MG-E)} framework.
Specifically, multiple generators are included to collaboratively produce valuable and diverse candidate lists before feeding the evaluator.
On the one hand, the combined generators form a list-level ensemble model which can easily avoid the potential local optimum of one single generator, offering more high-quality choices for the evaluator.
Consequently, the overall accuracy is theoretically and empirically superior to the single-generator methods.
On the other hand, restricting the list generation opportunities to the selected generators significantly reduces the number of lists to evaluate in the next evaluator step, compared to the permutation-based solutions.

Despite these advantages, the independent generators may have a chance of generating rankings that are too similar in practice (see $P_1$ and $P_2$ in Figure \ref{fig: intuition} for example), especially when they are pre-trained with the dataset under the same user environment .
This lack of list-level diversity would potentially reduce the effectiveness of the multi-generator design and we denote this problem as the \textbf{List Comprehensiveness Challenge}.
Note that this challenge does not exist in the permutation-based list exploration in the G-E framework, but rather a unique challenge under the MG-E framework.
To address this challenge, we further enhance the MG-E framework with a \textbf{List Comprehensiveness(LC) metric} that evaluates how different the generated lists are and guides the selection of generators.
We present an intuitive example in Figure \ref{fig: intuition}, and we can see that pairing the $P_1$ generator with $P_3$ would help increase the chance of finding the optimal list due to their diverse behaviors, while pairing $P_1$ with $P_2$ fails to achieve this goal.
Without loss of generality, all generators may have their own local optimal list covering one side of the ground-truth distribution, but we need a method to ensure their differences, which would benefit the ensemble effect of the generators.
We provide an empirical analysis in section \ref{sec: method_lc_metric} on how this metric correlates with the final reranking performance through different combinations of generators.

In addition, there are still limitations of manually designed generator combinations even with the LC metric as guidance:
1) The generator selection space grows combinatorially with $M$, making it increasingly difficult to analyze; 2) And there might still exist certain local optima patterns that no existing reranking solution can effectively cover.
Using the same example in Figure \ref{fig: intuition}, both $P_2$ and $P_3$ align with user preference, but $P_3$ has better LC performance with $P_1$ as the existing generator, thus the framework will automatically learn a generator that is close to $P_3$.
As a result, we derive a simple but effective learning framework that automatically finds a generator that complements the existing generators (See details in section \ref{sec: method_complementary_list_generation}).
During training, the corresponding model maximizes the gain of the overall LC metric, as well as the alignment with the user preference.
To illustrate the generalizability of the learning algorithm, we derive corresponding detailed solutions for both autoregressive and non-autoregressive list generation processes.
We denote this general extension technique as \textbf{Complementary LIst Generation (CLIG)} and verify its effectiveness in improving the ensemble effect through offline experiments in two public datasets, as well as online A/B test in our industrial recommender system.

In the remainder of this paper, we will present the detailed support for the following contribution:
\begin{itemize}[leftmargin=*]
    \item We prove that the MG-E framework is a superior design over generator-only solutions and perturbation-based G-E solutions.
    \item We identify the list comprehensiveness challenge in the MG-E framework and explain its relationship with the list-level ensemble effect through the LC metric.
    \item We propose the CLIG extension technique that can further boost the reranking performance and verify the sub-optimality of single-generator solutions and manually designed MG-E solutions.
\end{itemize}

\section{Related Work}\label{sec: related_work}

\subsection{Reranking Recommender System}

The notion of reranking in recommender systems has been around since the 90's~\cite{carbonell1998mmr,joachims2005interpret} where researchers have found that the item mutual influence (e.g. diversity) cannot be modeled by the standard learning-to-rank methodology~\cite{burges2005ltr,cao2007listwise}, and it has to advance the optimization problem into the list-wise viewpoint.
In other words, the reranking problem is generally formulated as a task that maximizes the list-wise reward given the entire initial list as input, assuming the existence of item mutual influences.
Following this setting, several generator-only approaches have been proposed to model and rerank the list as a whole~\cite{zhuang2018mutual,ai2018listwise,pei2019prm,gong2022edgererank,liu2023gfn4list}.
Some recent work~\cite{ren2024rlm4rec,gao2024llm} also noticed that large language models~\cite{wu2024survey} can enhance the reranking performance when textual information of items is provided.
However, effective and efficient exploration in the permutation space of the recommendation list is challenging, and the ground truth of a reranked list may not be presented in the original dataset, causing a lack of guidance.
As a solution to these limitations, the G-E framework has been proposed~\cite{shi2023pier,xi2024utility,lin2024dcdr,ren2024nar}.
On one hand, the evaluation of a list is a much simpler task than list generation, so making the evaluator a final list selector potentially improves the possibility of finding better rerankings.
On the other hand, the generator can effectively explore the permutation space through the guidance of the evaluator.
Our proposed method further improves this framework in the generator step, where a multi-generator design rather than a single-generator design is adopted to generate high-quality and diverse lists.

\subsection{List Generation in Recommendation}

There has been a research effort that studies the list-wise modeling in recommender systems.
Recently, the research focus has moved from discriminative methods that optimize a list-wise metric~\cite{cao2007listwise,xia2008listwise,burges2010ranknet,gong2019exactk,ie2019slateq} into generative methods~\cite{bello2018seq2slate,liu2021pivotcvae,liu2023gfn4list,lin2024dcdr} that model the probabilistic distribution of lists, which are further categorized into autoregressive methods and non-autoregressive methods.
The autoregressive approach~\cite{bello2018seq2slate,liu2023gfn4list} picks one item at a time during inference, which is favored in language-style scenarios (e.g., users sequentially browse the recommendation list).
While the generative model captures the next's relation to the previously selected items, its expected influence on ``future'' items is captured by the learning objective.
As a result, the majority of this type of solution is essentially an energy model that is simultaneously a generator and an evaluator.
For reranking tasks, existing works in~\cite{bello2018seq2slate,gong2022edgererank} fall in this category.
The non-autoregressive approach~\cite{liu2021pivotcvae,lin2024dcdr} directly modifies and outputs the list as a whole, which is well-suited for picture-style scenarios (e.g., presenting a list on one page).
While ~\cite{liu2021pivotcvae} generates all the items in the embedding space, ~\cite{lin2024dcdr} changes the entire list from the initial ranking with trivial actions like swapping.
For reranking tasks, existing works in~\cite{ren2024nar,lin2024dcdr} as well as our proposed framework belong to this category.

\subsection{Recommendation Diversity}

In addition to the ranking accuracy metrics, the diversity metric is also considered critical for recommendation in practice, and there has been a decent amount of research work on diversity modeling from two decades ago to today~\cite{carbonell1998mmr,ziegler2005diversity,cheng2017learning,chen2018dpp,sun2020framework,zheng2021dgcn,lin2022feature,xu2023multi,fu2023deep}.
The main application of diversity metrics are list-level metrics such as maximal marginal relevance~\cite{carbonell1998mmr}, intra-list diversity~\cite{ziegler2005diversity}
, $\alpha$-NDCG~\cite{clarke2008novelty}, and determinantal point process score~\cite{gartrell2016dpp,chen2018dpp}.
Other works also reveal that system-level diversity metrics such as item coverage~\cite{good1999coverage}, entropy~\cite{qin2013entropy}, GINI-index~\cite{ren2023slate}, and list variance~\cite{liu2021pivotcvae} are also important for maintaining a content-productive environment.
In this work, we address the importance of cross-list diversity between different list generators in the MG-E framework (i.e., the list comprehensiveness), which is close to the idea of list variance~\cite{liu2021pivotcvae}, but different in that the ordering of items is also considered and the metric is evaluated before an evaluator stage for better guidance of list generation.

\section{Method}\label{sec: method}

\subsection{Problem Formulation}\label{sec: method_problem_formulation}

\begin{figure}[t]
    \centering
    \begin{subfigure}[b]{\linewidth}
        \centering
        \includegraphics[width=0.85\linewidth]{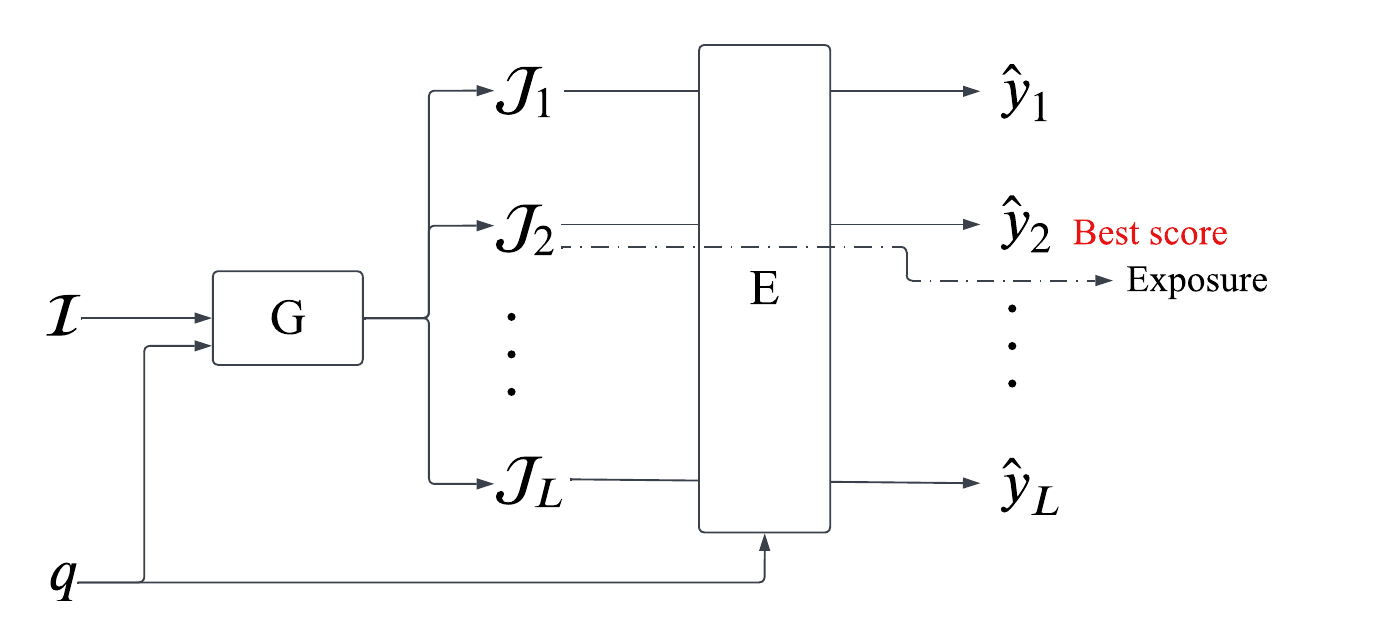}
        \caption{Standard generator-evaluator framework for reranking}
    \end{subfigure}
    \begin{subfigure}[b]{\linewidth}
        \centering
        \includegraphics[width=0.85\linewidth]{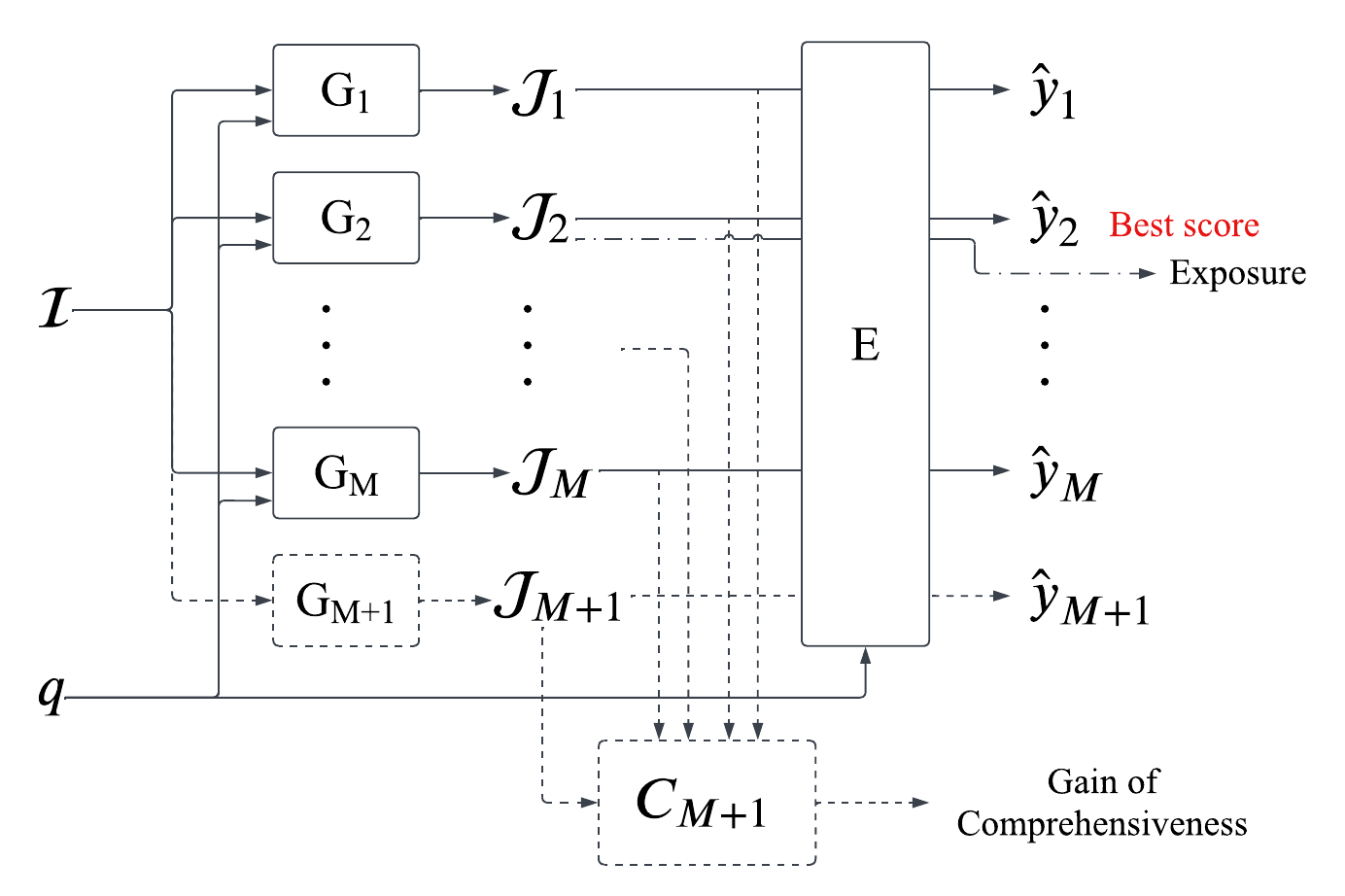}
        \caption{The proposed multi-generator-evaluator framework with list comprehensiveness evaluation}
    \end{subfigure}
    \caption{Comparison of overviews between standard G-E and the proposed MG-E framework with CLIG extension.}
    \label{fig: overview}
\end{figure}

A user request $q\in\mathbb{R}^a$ in the recommendation service encodes the information about the user's current state, which includes the recent interaction history that describes the dynamic preference transition and the static profile features (such as user ID, gender, and location) that identify collective biases in the preference.
In the final reranking stage, an ordered list of items $\mathcal{I}=[i_1,i_2,\cdots,i_n]$ is pre-selected by previous stages in the multi-stage recommendation process.
The corresponding reranking (i.e., the generator) model $G(q,\mathcal{I})$ takes the user request and the candidate list of items as input and returns a reranked list of items $\mathcal{J}=[j_1,j_2,\cdots,j_n]$ as output.
The learning objective is defined as the maximization of the user feedback (i.e., $\max_\theta \mathcal{R}(q,\mathcal{J})$) through reranking for any given user request $q$,
where the reward model could adopt any listwise metric like the aggregation of positive feedback~\cite{ren2024nar} and relevance ranking metrics~\cite{ai2018listwise}.
During inference of a standard generator-evaluator (G-E) framework~\cite{shi2023pier,ren2024nar,lin2024dcdr}, the evaluator will estimate all the lists generated by the generator and select the list with the best score as the final recommendation (i.e., Figure \ref{fig: overview}-a).

\subsection{Multi-Generator Reranking with Evaluator}\label{sec: method_mg_e}

Our proposed MG-E framework first distinguishes itself from standard G-E by the multi-generator design.
As shown in Figure \ref{fig: overview}-b, the evaluator behaves in the same way as the G-E framework, but $M$ different generator models $G_1,\cdots,G_M$ are included (excluding the $(M+1)$-th extension) and each generator will propose a reranked list to the evaluator.
During training, all generators maintain their own training objectives along with the training of the evaluator.
In this work, we focus on the design in the generator step and consider the study of the evaluator step as complementary.
To keep a fair comparison, we fix the evaluator architecture as the same as the OCPM model in PIER~\cite{shi2023pier} since it has the state-of-the-art ability to model mutual influence between items in a list.

\subsubsection{Superiority:} In a probabilistic view, each generator $G_m$ determines the generation probability $P(\mathcal{J}|q,\mathcal{I})$ for any permutation $\mathcal{J}$.
For deterministic models, only the greedy output list $\mathcal{J}_m$ has probability $P(\mathcal{J}_m|q,\mathcal{I})=1$, and all other list permutations assign zero probability.
Theoretically, this design forms an ensemble~\cite{opitz1999ensemble,mienye2022survey} of reranking models, where the overall performance is improved as the number of models increases, for the probability of selecting the optimal list $J_\ast$ is boosted monotonically with $M$:
\begin{equation}
    P(\mathcal{J}_\ast|q,\mathcal{I}) = 1 - \prod_{m=1}^M (1-P_m(\mathcal{J}_\ast|q,\mathcal{I})),
\end{equation}
where the probability $P_m(\mathcal{J}_\ast|\cdot)$ denotes the probability of generating the optimal list from the certain generator $G_m$.
We can also derive a similar conclusion for the $b$ most optimal lists by substituting $P_m(\mathcal{J}_\ast|q,\mathcal{I})$ with $P_m(\cdot|q,\mathcal{I})$.

\subsubsection{Exploration efficiency:} Note that the standard G-E framework~\cite{shi2023pier} allows the lone generator to produce more than one list through random permutation, and increasing the number of lists improves the chance of hitting the optimal list.
As we have mentioned in section \ref{sec: introduction}, this naive permutation-based list generation is not a cost-efficient exploration strategy under the exponential list permutation space.
Specifically, the permutation-based method is effective in finding ``similar and high-valued'' lists around the local optima, but becomes inefficient in finding ``high-valued but strikingly different'' lists.
In contrast, the multi-generator design is a superior design in solving this limitation by breaking the local optimum restriction (as empirically proved in other generative tasks~\cite{khayatkhoei2018disconnected,hoang2018mgan}), and thus improves the list exploration efficiency.
This also provides an analytical foundation for our designs of the list comprehensive metric and the complementary list generation technique in the following sections.

\subsection{List Comprehensiveness}\label{sec: method_lc_metric}

A good ensemble is one where the individual models are both accurate and behaviorally different~\cite{opitz1999ensemble}.
Yet, the straightforward MG-E framework guarantees the generation of accurate lists but not necessarily the list diversity, which potentially reduces the effectiveness of the multi-generator ensemble.
As we have illustrated in Figure \ref{fig: intuition} and section \ref{sec: introduction}, generators may still output the same list for a given $q$ and $\mathcal{I}$, especially when they are trained from the same dataset.
We denote this distinctive issue of the MG-E framework as the \textbf{list comprehensiveness challenge}.

\subsubsection{Metric definition:}
To better understand this challenge, we first address the importance of the evaluation of the list-level diversity and observe the following two metrics:
\begin{itemize}[leftmargin=*]
    \item \textbf{List comprehensiveness} $\text{LC}_m$ evaluates how a given set of lists $\mathcal{J}_1,\cdots,\mathcal{J}_m$ are different from each other:
    \begin{equation}
         \text{LC}_m = \sum_{m_1,m_2 < m, m_1\neq m_2} \delta(\mathcal{J}_{m_1},\mathcal{J}_{m_2}),\label{eq: lc_metric}
    \end{equation}
    where $\delta$ denotes the pairwise diversity evaluation kernel. Intuitively, one can improve this metric either by adding a new list or by substituting a list that is more different from other lists.
    \item \textbf{Gain of Comprehensiveness} $\text{GC}_m$ evaluates how much a newly added generator $G_m$ contributes to the LC metric:
    \begin{equation}
        \text{GC}_m=\text{LC}_m - \text{LC}_{m-1}=\sum_{m^\prime=1}^{m-1}\delta(\mathcal{J}_{m^\prime},\mathcal{J}_{m}).\label{eq: gc_metric}
    \end{equation}
\end{itemize}
Note that these metrics estimate cross-list diversity rather than intra-list diversity, so the kernel function $\delta$ should represent how different the two given rerankings are.

\subsubsection{Choices of list diversity kernel:}\label{sec: methods_diversity_kernel} In practice, when the item embedding is available, one can adopt \textbf{embedding-based diversity (EBD)} estimation:
\begin{equation}
    \delta(\mathcal{J}_m,\mathcal{J}_{m^\prime}) = 0.5\big(1 - \cos (E(\mathcal{J}_m),E(\mathcal{J}_{m^\prime}))\big) ,
\end{equation}
where the embedding function $E(\cdot)$ may use position-weighted pooling of item embeddings when the input is a list, and each item embedding is extracted either through the initial ranker or other pretrained content encoders~\cite{zhang2024notellm,chen2024m3}.
Note that we can also regard each list as a cluster and the kernel as cluster diversity measurement, then existing methods like minimal linkage, maximal linkage, and centroid linkage~\cite{xu2005survey} are also feasible.
Yet, in cases where the embedding is not available or not in the same unified space across generators, one can adopt a more general kernel function that calculates the total \textbf{partial order disparity (POD)}:
\begin{equation}
    \delta(\mathcal{J}_m,\mathcal{J}_{m^\prime}) = \sum_{\substack{i,j\in\mathcal{J}_m \\ r(i,\mathcal{J}_m)<r(j,\mathcal{J}_m)}}\mathbb{I}[r(i,\mathcal{J}_{m^\prime})>r(j,\mathcal{J}_{m^\prime})],
\end{equation}
where $r(\cdot,\mathcal{J})$ denotes the rank of an item in the ordered list $\mathcal{J}$, and the disparity is larger when the two lists have more disagreement in the partial orders.
The metric is minimized when the two lists are identical and maximized when one list is the reverse of the other.
Compared to EBD, the POD metric is independent of the embeddings and generator models, but it takes more effort to calculate and is less fine-grained than embedding-based distance measures.

We remind readers that, in addition to the standard $n\rightarrow n$ reranking task, there are other reranking scenarios~\cite{ren2024nar} that include an extra top-$k$ ($k<n$) filtering step after generation.
In these circumstances, we can only use POD for the lists (each with size $n$) before the filtering step, but not the post-filtering lists (each with size $k$) since the items from different lists may no longer be the same.
In contrast, the EBD kernel can effectively estimate both the pre-filtering list (of size $n$) and the post-filtering list (of size $k$).
We summarize the characteristics of the two kernels in Table \ref{tab: kernel_comparison}.

\begin{table}[t]
    \centering
    \caption{Qualitative comparison between EBD and POD diversity kernels for LC metrics.}
    \begin{tabular}{c|c|c|c}
        \toprule
        kernel & allow $n\rightarrow k$ filtering & model-agnostic & granularity \\
        \midrule
        EBD & Yes & No & Fine\\
        POD & No & Yes & Coarse\\
        \bottomrule
    \end{tabular}
    \label{tab: kernel_comparison}
\end{table}

\subsubsection{Empirical study:}\label{sec: lc_metric_empirical_study} To further investigate the connection between the LC metric and the ensemble performance, we conduct an offline experiment on the Avito dataset\footnote{https://www.kaggle.com/c/avito-context-ad-clicks/data} and compare different generator combination strategies.
Specifically, we share an OCPM evaluator~\cite{shi2023pier} in the MG-E framework and pick three generator candidates: G1~\cite{covington2016deep}, an item-wise neural model; G2~\cite{ren2024nar}, a permutation-based non-autoregressive model; G3~\cite{bello2018seq2slate}, an auto-regressive generative model.
We set $n=5$ for the list size and the permutation space has size $n!=120$.
We summarize the 5-round (random-seeded) average results in Table \ref{tab: lc_performance}.
We can see that increasing the number of generators $M$ generally improves the recommendation performance (i.e., AUC) along with the LC measure, and the three-generator method (i.e., G1+G2+G3) can more effectively boost the performance compared to PIER~\cite{shi2023pier} that yields more than three lists.
This verifies the effectiveness of the multi-generator ensemble with different types of generators.
Among MG-E methods with two generators, the AUC metric is also positively related to the LC and GC metrics, indicating the importance of choosing generators that provide diverse and comprehensive candidate lists.
Yet, this difference also means that a good manual design requires expert knowledge in practice, and we would like to automatically find a generator that can fulfill this task.
\begin{table}[t]
    \centering
    \caption{Performance comparison between different generator combinations on Avito dataset.}
    \begin{tabular}{c|c|c|c|c|c}
        \toprule
        framework & generators & AUC & LC(POD)  & LC(EBD)  & \#list\\
        \midrule
        \multirow{2}{*}{G-only} & G1 & 0.6876  & - & -  & 1 \\
        & G3 & 0.6911 & -  & -  & 1\\
        \midrule
        \multirow{3}{*}{G-E}& G2 & 0.7066  & - & - & 1\\
        & PIER-3 & 0.7180 & 9.9042  & 0.2913  & 3\\
        & PIER-20 & 0.7185 & 313.5678 & 0.6481  & 20\\
        \midrule
        \multirow{5}{*}{MG-E} & G1+G2  & 0.7113 & 8.1562  & 0.4239  & 2\\
        & G1+G3 & 0.7121 & 9.8750  & 0.7398 & 2\\
        & G1+CLIG & 0.7135 & 10.0312  & 0.7681  & 2\\
        \cline{2-6}
        & G1+G2+G3  & 0.7225 & 10.2187  & 0.7884  & 3\\
        & +CLIG & 0.7297 & 11.3437  & 0.8722  & 4\\
        \bottomrule
    \end{tabular}
    \label{tab: lc_performance}
\end{table}
\subsection{Learning Complementary List Generation}\label{sec: method_complementary_list_generation}

As we have illustrated in section \ref{sec: introduction}, to mitigate the deficiencies of the manually designed MG-E framework, we propose the CLIG extension technique that automatically learns a complementary list generator that can simultaneously align with user preferences and improve the LC metric over the existing generators.

\subsubsection{Methodology:} Formally, we first assume that there are $M$ existing generators (i.e., $G_1,G_2$ $,\cdots,G_M$) in the MG-E framework, then aim to include a new generator $G_{M+1}$ (with parameter $\theta$) that can simultaneously improve the LC metric and align the reranking with the user preferences, following the notion of finding ``high-valued but different'' lists.
We assume an item-separable reward model~\cite{ren2024nar} (i.e., $\mathcal{R}(q,J) = \sum_{j\in\mathcal{J}}\mathcal{R}(q,j)$).
The new generator $G_{M+1}$ determines the list generation probability $P_\theta(\mathcal{J}_{M+1}|q,\mathcal{I})$.
The corresponding learning objective for the user preference alignment uses a reward-weighted cross-entropy loss for a generated $\mathcal{J}_{M+1}\sim G_{M+1}(q,\mathcal{I})$:
\begin{equation}
    \mathcal{L}_\text{align} = -\mathcal{R}(q,\mathcal{J}_{M+1}) \log P_\theta(\mathcal{J}_{M+1}|q,\mathcal{I})
\end{equation}
where lists with better rewards tend to further increase the generation probability.
For auto-regressive models, we can assign each item-wise reward to the step-wise probability:
\begin{equation}
    \mathcal{L}_\text{align\_ar} = -\sum_{t=1}^n\mathcal{R}(q,j_t) \log P_\theta(j_t|q,\mathcal{I},j_{0:t-1})
\end{equation}
where the auto-regressive model takes the previously selected items $j_{0:t-1}$ as an extra input and $j_0$ denotes the initial empty list.

Besides, we include the list comprehensiveness enforcement loss that directly uses the GC metric in Eq.\eqref{eq: gc_metric}:
\begin{equation}
\begin{aligned}
    \mathcal{L}_\text{comp} & = -\text{GC}_{M+1}\log P_\theta(\mathcal{J}_{M+1}|q,\mathcal{I}) \\
    & = -\Big(\sum_{m^\prime=1}^M \delta(\mathcal{J}_{m^\prime},\mathcal{J}_{M+1})\Big) \log P_\theta(\mathcal{J}_{M+1}|q,\mathcal{I})\label{eq: comp}
\end{aligned}
\end{equation}
which encourages the generation of lists with larger gains of comprehensiveness.
For autoregressive generators, we adopt the item-wise GC in the modified loss:
\begin{equation}
    \mathcal{L}_\text{comp\_ar} = -\sum_{t=1}^n\Big(\sum_{m^\prime=1}^M \delta(\mathcal{J}_{m^\prime},j_t)\Big) \log P_\theta(j_t|q,\mathcal{I},j_{0:t-1})\label{eq: comp_ar}
\end{equation}
where the diversity kernel $\delta$ estimates the difference between an item and a list.
In this case, the EBD kernel also makes the corresponding substitution with item embedding, while the POD kernel only considers the partial ordering related to the given item $j_t$.
We provide a detailed description of this alternative in Appendix \ref{ap: autoregressive_kernel}.

During training, we combine the two objectives as the final loss function and adopt stochastic gradient descent optimization on $\theta$:
\begin{equation}
    \mathcal{L} = \mathcal{L}_\text{align} + \lambda \mathcal{L}_\text{comp},\label{eq: total_clig_loss}
\end{equation}
where the $\lambda$ coefficient controls the importance of list comprehensiveness boosting, and the same strategy also applies to the autoregressive models.
In the remainder of the paper, we assume the autoregressive model implementation for CLIG due to its superior alignment ability for the sequential recommendation task.

\subsubsection{Empirical Study:} Using the same experimental setting in Table \ref{tab: lc_performance}, we compare the two-generator methods with G1 as the existing generator.
Compared to G2 and G3, the CLIG extension can more effectively improve the LC metric and further improve the recommendation accuracy compared to manually selected generators.
Additionally, according to~\cite{liu2021pivotcvae}, boosting the cross-list variance (through the LC metric) theoretically guarantees a lower bound on the overall recommendation diversity.
We will also provide empirical proof for this connection in section \ref{sec: experiments_online}.

\subsubsection{Including Multiple Extensions.}\label{sec: method_multiple_extensions}
Note that the CLIG technique is an incremental strategy that does not restrict the choice of backbone generators, so the CLIG extension itself can be considered as the backbone and iteratively increases the number of extensions.
In section \ref{sec: experiments_offline}, we will show that CLIG is a general technique that is effective on different base rerankers, and it can further improve the performance by including more extensions.
Intuitively, the combination of different CLIG extensions forms an ensemble that tries to behave differently from each other, similar to the principal component decomposition~\cite{wold1987principal,kherif2020principal} of data variance.

\subsection{Computational Cost and Latency}\label{sec: method_computational_cost}

Compared to single-generator methods, the MG-E framework, either using manual design or CLIG technique, linearly increases the computational cost with the number of generators $M$.
One can avoid the latency overhead by running all generators in parallel, but the computational cost is inevitable.
As the number of generators increases, the chance of finding high-value but different lists decreases, so the marginal improvement in recommendation performance theoretically shrinks towards zero as $M$ increases:
\begin{equation}
\begin{aligned}
    & \lim_{M\rightarrow \infty} \big(P_{1:M+1}(\mathcal{J}_\ast|q,\mathcal{I})-P_{1:M}(\mathcal{J}_\ast|q,\mathcal{I})\big) \\
    = & \lim_{M\rightarrow\infty} P_{M+1}(\mathcal{J}_\ast|q,\mathcal{I})\prod_{m=1}^M(1- P_{m}(\mathcal{J}_\ast|q,\mathcal{I}))= 0,
\end{aligned}
\end{equation}
Thus, the designer has to balance between the marginally decreasing gain of utility and the linearly increased computational cost in practice.
Though not the focus of our paper, we believe that there exists a minimal $M$ that achieves the best balance, and the CLIG extension technique potentially approximates the optimal solution through a greedy generator learning method.

\section{Experiments}\label{sec: experiments}

We summarize the key research questions that require empirical support as follows:
\begin{itemize}[leftmargin=*]
    \item \textbf{RQ1:} Proof of MG-E's superiority in the reranking task, compared with single-generator methods.
    \item \textbf{RQ2:} Proof of the effectiveness of CLIG extension technique that further boosts the MG-E performance.
    \item \textbf{RQ3:} Investigation of LC metrics and GC's effectiveness under different hyperparameters including $\lambda$, choices of kernel functions, and the number of extensions.
\end{itemize}

\subsection{Offline Experiments}\label{sec: experiments_offline}

\subsubsection{Experimental Setup}

\begin{table}[]
    \centering
    \caption{Statistics of public datasets in offline experiments}
    \begin{tabular}{c|c|c|c|c}
        \toprule
        datasets & \#items & \#users & \#requests & positive rate\\
        \midrule
        Avito & 23,562,269 & 1,324,103 & 53,562,269 & 3.516\%\\
        RecFlow & 1,361,856 & 652,490 & 24,637,522 & 60.656\%\\
        \bottomrule
    \end{tabular}
    \label{tab: public_datasets}
\end{table}
\begin{itemize}[leftmargin=*]
\item \textbf{Dataset}: In our experiments, we include two public datasets, Avito and RecFlow. 
The Avito dataset is a user search log dataset that contains tens of millions of search requests along with the recommendation list and user click responses with $n=5$, and we follow similar preprocessing steps as PIER~\cite{shi2023pier}. 
The RecFlow dataset~\cite{liu2025recflow} is a recently published multi-stage video recommendation dataset where the reranking task is included as a sub-problem.
We adopt an 80-20 split on the dataset according to the timestamp, and for each user's interaction history (with effective\_view = 1), we segment it into lists of size $n=6$ and consider long\_view = 1 (i.e., watching sufficiently long time of a video) as positive signals.
We summarize the statistics of the two preprocessed datasets in Table \ref{tab: public_datasets}.
\item \textbf{Evaluation protocol}: We assume a $n\rightarrow n$ reranking scenario and evaluate the AUC and NDCG of the reranked list with user feedback as ground-truth relevance signals.
Both metrics are higher when items with higher rewards are reranked in the front of the resulting list.
Additionally, to better investigate the correlation between the cross-list diversity and the final recommendation performance, we also include the LC and GC metrics introduced in section \ref{sec: method_lc_metric}.
\item \textbf{Baselines}: We include several types of baseline models including 
a) the item-wise score estimators DNN~\cite{covington2016deep} and DCN~\cite{wang2017dcn} that separately learn the user feedback for each user-item interaction; 
b) Seq2Slate~\cite{bello2018seq2slate} that uses pointer-network to learn the ordering of lists, which is an auto-regressive generator-only solution;
c) PRM~\cite{pei2019prm} and EXTR~\cite{chen2022extr} that use transformer architecture to model the mutual influences between items in the list, where we consider them as representatives of non-autoregressive generator-only solutions;
d) PRS~\cite{feng2021prs}, PIER~\cite{shi2023pier}, and NAR4Rec~\cite{ren2024nar} that adopt generator-evaluator framework with single-generator in the first step.
Note that the state-of-the-art method PIER allows the single generator to generate multiple lists through permutation-based operations, so we specify the PIER-$L$ alternatives where $L$ represents the number of generated lists.
For PIER, while setting the maximum value of $L=n!$ is impractical, we restrict it to the cases with $L=3$ (compared with the MG-E base method) and $L=20$ (the original setting in PIER).
The NAR4Rec baseline also mentioned the possibility of adding its generator inside an MG-E framework but did not publish the design, so we replicated its implementation according to the paper and considered it as a G-E method.
\end{itemize}

For all methods, we fix the optimizer to Adam and grid search the learning rate in [0.00001, 0.0001, 0.001, 0.01], the mini-batch size in [256, 512, 1024].
We pick the setting with the best test result as offline validation for later online experiments.

\subsubsection{Model Specification:} For the proposed MG-E framework, we use the three-generator design mentioned in section \ref{sec: lc_metric_empirical_study} as our base method, which uses an ensemble of DNN, NAR4Rec, and Seq2Slate in the generator step, and OCPM~\cite{shi2023pier} in the evaluator step.
During training, we first train the evaluator until convergence, then we separately train each generator.
For neural network design, we suggest a transformer-based architecture~\cite{vasvani2017attention,pei2019prm} to better capture the item mutual influences among the items in the initial list and integrate an extra GRU~\cite{balazs2016gru4rec} layer if using the autoregressive framework, similar to our model design in the online experiments shown in Figure \ref{fig: online_workflow}.
In practice, we have not found a significant difference between different architectures of reranking models in terms of the final recommendation performance as long as the transformer architecture is involved, so we keep our solution description in a model-agnostic manner and focus on the analysis of the learning paradigm.
We provide the source code\footnote{\url{https://github.com/hellohelen222/LC_MGE_code}} for reproduction.

\subsubsection{Main Results (RQ1)}\label{sec: experiments_offline_main_results}

\begin{table}[t]
    \centering
    \caption{Recommendation performance comparison in offline experiments. Values in bold represent best results and values with underline represent second best results.}
    \begin{tabular}{c|c|c|c|c|c}
        \toprule
        \multirow{2}{*}{framework} & \multirow{2}{*}{method}
        & \multicolumn{2}{c}{Avito} & 
        \multicolumn{2}{c}{RecFlow} \\
        & & AUC & NDCG & AUC & NDCG \\
        \midrule
        \multirow{5}{*}{G-only} & DNN & 0.6876 & 0.6028& 0.7067 & 0.6831\\
        & DCN & 0.6896 & 0.6059 & 0.7088 & 0.6856 \\
        & Seq2Slate & 0.6911 & 0.6115 & 0.7098& 0.6898 \\
        & PRM & 0.7124 & 0.6245 & 0.7613 & 0.7004 \\
        & EXTR & 0.7113 & 0.6228 & 0.7641 & 0.7082 \\
        \midrule
        \multirow{4}{*}{G-E} & NAR4Rec & 0.7066 & 0.6205 & 0.7284 & 0.6979\\
        & PRS & 0.7182 & 0.6301 & 0.7639 & 0.7067\\
        & PIER-3 & 0.7180 & 0.6273 & 0.7640 & 0.7012\\
        & PIER-20 & 0.7185 & 0.6327 & 0.7645 & 0.7102 \\
        \midrule
        \multirow{2}{*}{MG-E} & Base & \underline{0.7225} & \underline{0.6487} & \underline{0.7842} & \underline{0.7217} \\
        & +CLIG & \textbf{0.7297} & \textbf{0.6629} & \textbf{0.7878} & \textbf{0.7357} \\
        \bottomrule
    \end{tabular}
    \label{tab: main_result}
\end{table}

We conduct each experiment for five rounds with different random seeds and report the result in table $\ref{tab: main_result}$.
On both datasets, the state-of-the-art G-E methods generally achieve better AUC and NDCG, compared with the generator-only methods, except that the EXTR baseline appears to be better than the (restricted) NAR4Rec baseline and only slightly worse than PIER.
In contrast, the proposed MG-E solutions consistently outperform all methods, illustrating the ensemble effect.
When generating the same number of lists (i.e. $M=L=3$), compared with PIER-3, the base model of MG-E improves the AUC by 0.5\% on Avito and 2.6\% (with student-t significance test value of $p<0.005$).
Additionally, the PIER-20 model increases the number of lists to 20 but only marginally improves the reranking accuracy, while the MG-E continues to be superior to this alternative.
These results indicate that the carefully designed MG-E solution can better explore the list permutation space and find better reranking.

\subsubsection{Effectiveness of CLIG Extension (RQ2)}\label{sec: experiments_clig_effectiveness}

\begin{table}[t]
    \centering
    \caption{The comparison result between G-E backbones and their CLIG-extended counterparts.}
    \begin{tabular}{c|c|c|c|c}
        \toprule
        \multirow{2}{*}{method} & \multicolumn{2}{c}{Avito} & \multicolumn{2}{c}{RecFlow} \\
        & AUC & NDCG & AUC & NDCG \\
        \midrule
        NAR4Rec & 0.7066 & 0.6205 & 0.7284 & 0.6979 \\
        NAR4Rec + CLIG & 0.7154 & 0.6269 & 0.7375 & 0.7011 \\
        Improvement & \textbf{+1.259\%} & \textbf{+1.031\%} & \textbf{+1.249\%} & \textbf{0.458\%} \\
        \midrule
        PRS & 0.7182 & 0.6301 & 0.7639 & 0.7067\\
        PRS + CLIG & 0.7198 & 0.6350 & 0.7698 & 0.7112\\
        Improvement & \textbf{+0.236\%} & \textbf{+0.777\%} & \textbf{+0.772\%} & \textbf{+0.637\%} \\
        \midrule
        PIER-20 & 0.7185 & 0.6327 & 0.7645 &0.7102\\
        PIER-20 + CLIG & 0.7211 & 0.6468 & 0.7735 & 0.7221\\
        Improvement & \textbf{+0.362\%} & \textbf{+2.228\%} & \textbf{+1.177\%} & \textbf{+1.675\%} \\
        \midrule
        MG-E & 0.7225 & 0.6487 & 0.7842 & 0.7217\\
        MG-E + CLIG & 0.7297 & 0.6629 & 0.7878 & 0.7357\\
        Improvement & \textbf{+0.927\%} & \textbf{+2.189\%} & \textbf{+0.459\%} & \textbf{+1.926\%} \\
        \bottomrule
    \end{tabular}
    \label{tab: clig_extension}
\end{table}

As shown in Table \ref{tab: main_result}, we can see that the CLIG extension technique can effectively improve the base MG-E model (also statistically significant).
This verifies that the GC-based guidance can effectively find the complementary generator and further improve the recommendation performance.
To verify the backbone-agnostic nature of CLIG, we integrate the CLIG extension to different G-E backbones and report the results in table \ref{tab: clig_extension}.
We observe that, for all tested backbones, adding the CLIG extension consistently improves the results, verifying its generalizability and flexibility.

\subsubsection{Ablation Study (RQ3)}

\begin{figure}
    \centering
    \includegraphics[width=\linewidth]{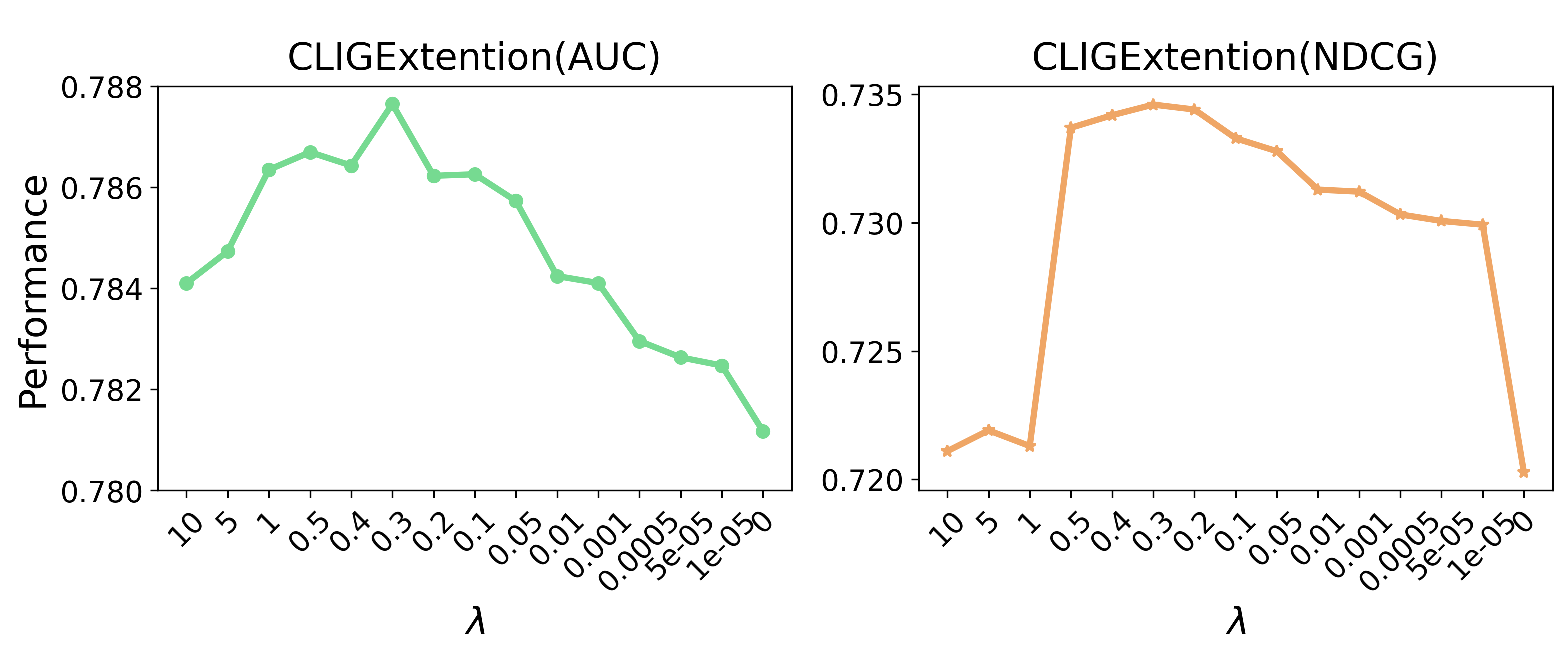}
    \caption{The comparison result of different settings of $\lambda$ for CLIG extension on MG-E base model.}
    \label{fig: lamba_auc_ndcg}
\end{figure}

\begin{itemize}[leftmargin=*]
    \item \textbf{Kernel functions:} We compare the two LC kernel functions (i.e., EBD and POD) mentioned in section \ref{sec: methods_diversity_kernel}, and report the result in table \ref{tab: kernel_comparison_empirical}.
    We can see that both alternatives improve the recommendation results as long as they describe the list-wise diversity, while the fine-grained guidance with EBD is slightly better than the coarse guidance with POD.
    \item \textbf{Magnitude of $\mathcal{L}_\text{comp}$:} 
    We conduct experiments with $\lambda\in[0, 10.0]$ to control the magnitude of GC reward in Eq.\eqref{eq: total_clig_loss} when accommodating CLIG on the MG-E base model.
    We present the results in Figure \ref{fig: lamba_auc_ndcg} and show that there exists an optimal point in $\lambda^\ast\in [0.1,0.5]$.
    Intuitively, restricting the $\lambda$ towards zero would make the extension close to learning an independent generator, which may not guarantee sufficiently diverse list exploration.
    On the other hand, setting $\lambda$ to a large value would overwhelm the learning of the original $\mathcal{L}_\text{align}$, which negatively impacts the recommendation accuracy.
    Neither extreme cases achieve better results than the intermediate optimal point.
    \item \textbf{The number of extensions:} 
    As we have illustrated in section \ref{sec: method_multiple_extensions}, we can iteratively use CLIG to increase the number of generators.
    We conduct a corresponding experiment that tests different numbers of iterative CLIG extensions and report the result in table \ref{tab: extension_number}, which verifies the incremental improvements of including more CLIG extensions.
    
\end{itemize}

\begin{table}[t]
    \centering
    \caption{Comparison results of EBD and POD kernels when adding the CLIG extension on MG-E base model.}
    \begin{tabular}{c|c|c}
        \toprule
        method & CLIG w/ EBD & CLIG w/ POD \\
        \midrule
        AUC & 0.7878 & 0.7863 \\
        Improvement & \textbf{+0.459\%} &  \textbf{+0.268\%} \\
        \bottomrule
    \end{tabular}
    \label{tab: kernel_comparison_empirical}
\end{table}

\begin{table}[t]
    \centering
    \caption{The comparison result of different number of CLIG extension on MG-E base model.}
    \begin{tabular}{c|c|c|c}
        \toprule
        \#CLIG Extension & 1 & 3 & 5 \\
        \midrule
        AUC & 0.7878 & 0.7882 & 0.7894 \\
        Improvement & \textbf{+0.459\%}& \textbf{+0.510\%}& \textbf{+0.663\%}\\
        \bottomrule
    \end{tabular}
    \label{tab: extension_number}
\end{table}

\subsection{Online A/B Test}\label{sec: experiments_online}

\begin{figure}[t]
    \centering
    \includegraphics[width=\linewidth]{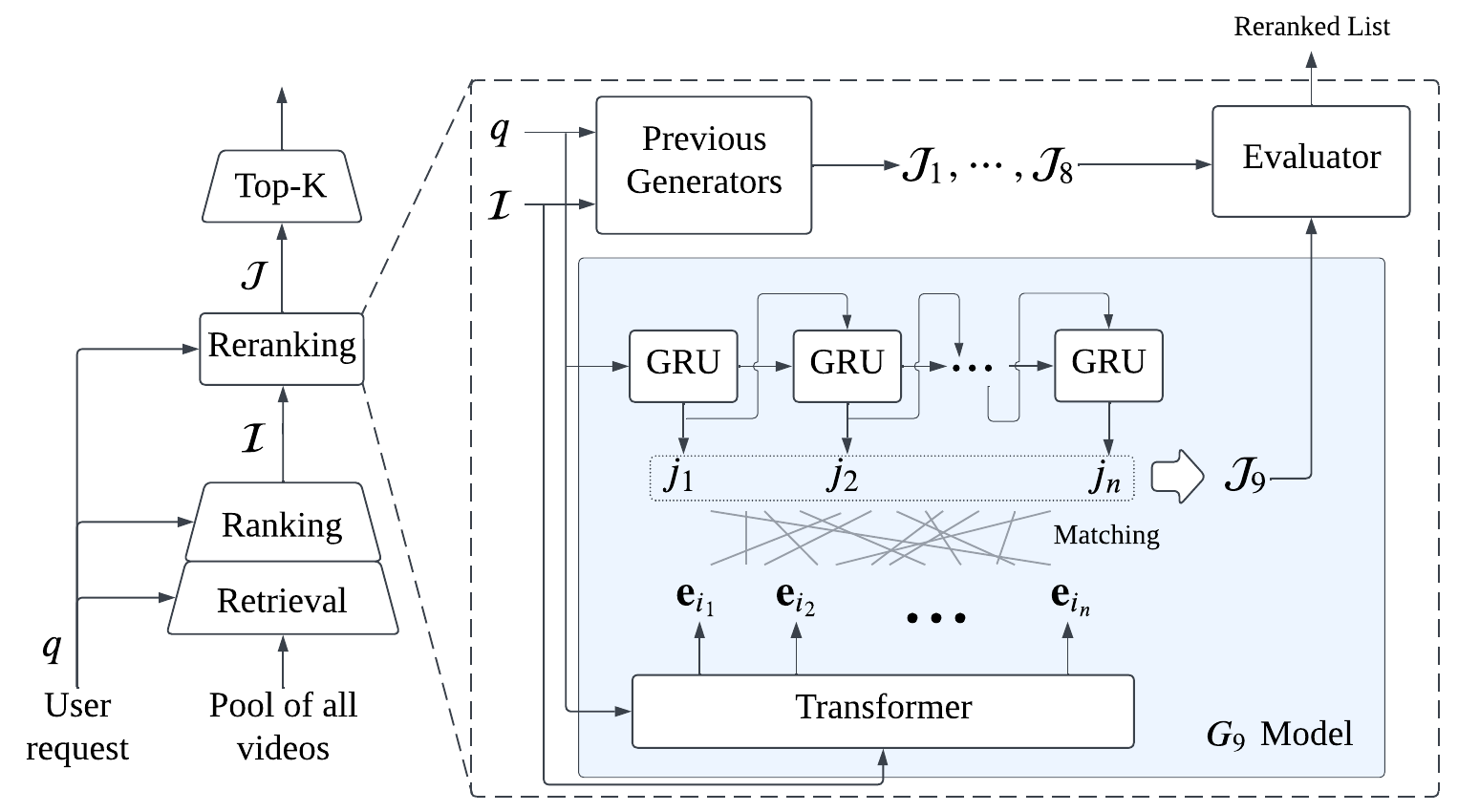}
    \caption{Online recommendation workflow and the detailed forward model of the newly added generator in CLIG.}
    \label{fig: online_workflow}
\end{figure}

To better investigate how the MG-E framework and the CLIG extension behave in an online environment, we conduct an A/B test on an industrial platform that serves video feeds to hundreds of millions of users on a daily basis.
As illustrated in Figure \ref{fig: online_workflow}, the general workflow of the online system is a multi-stage framework that consists of three stages: retrieval, ranking, and reranking.
The last stage conducts a top-$k$ selection after the $n\rightarrow n$ reranking process, where $n=50$ and $k=6$.
The permutation space near $1.5\times 10^{10}$, which is impractical to enumerate.

\subsubsection{Experimental Setup}
In our experimental design, the offline evaluation serves as the validation set to choose optimal hyperparameters, while the online A/B testing results are treated as the test set. 
In turn, the evaluation in the online environment with controlled A/B testing serves as the unbiased test set for reporting performance. 
This setup ensures a strict separation between validation and test sets, preventing information leakage and overfitting issues during evaluation.
\begin{itemize}[leftmargin=*]
    \item \textbf{Baselines and models}: we use PIER~\cite{shi2023pier} (with $L=50$) as the baseline and compare it with the MG-E base framework with $M=3$ manually designed models (which aligns with our setting in the offline experiments).
    Upon the time we test the CLIG extension, the MG-E framework has progressed to an eight-generator version ($M=8$).
    We omit the details of the previous $M$ generators and present a specification of the newly added $G_9$ model in the EBD-based CLIG extension, as shown in Figure \ref{fig: online_workflow}.
    For hyperparameters of CLIG, we set the kernel function to EBD, $\lambda=0.1$, and restrict to one CLIG extension.
    \item \textbf{Data stream}: For each experiment, we evenly separate the data traffic into ten folds, where two folds (taking 20\% of the data stream) are compared as the baseline and we deploy our model alternatives (i.e., MG-E and MG-E with CLIG) in 4 other folds, each taking 2 folds.
    The data samples are similar to that in the RecFlow dataset, where the users provide various types of feedback including but not limited to watch time, effective view, click like button, sharing, and following the author.
    \item \textbf{Training procedure}: All the aforementioned signals are linearly combined with empirical weights to obtain the item-wise reward during training.
    Since we adopt an autoregressive model in the CLIG extension, the corresponding learning objective becomes $\mathcal{L}=\mathcal{L}_\text{align\_ar} + \lambda \mathcal{L}_\text{comp\_ar}$.
    In online learning, training data continuously evolves along with the model.
    To avoid catastrophic forgetting~\cite{robins1995catastrophic,wang2024continual}, we adopt the replay buffer technique that stores recent samples and gives each historical sample a chance of being included in the training at any time.
    \item \textbf{Evaluation protocol}: for accuracy metrics, we directly use the key components in the reward, including the average watch time, effective view count, like count, share count, and follow count.
    Additionally, to verify the correlation between the CLIG extension and the traditional diversity metrics, we also observe several diversity metrics including category coverage, item coverage, and intra-list diversity~\cite{ziegler2005diversity}.
\end{itemize}

\subsubsection{Main Result (RQ1 and RQ2)}

For both the MG-E validation and the CLIG validation, we conduct corresponding experiments for one week and report the average result in Table \ref{tab: online_result_mg_e} and Table \ref{tab: online_result_clig}, respectively.
We can see that MG-E outperforms PIER in the reranking accuracy metrics but not necessarily in diversity metrics.
Then, adding the CLIG extension can automatically find a complementary generator that significantly improves the diversity and some of the key interaction metrics.
This also verifies the limitation of manually designed generators and the generalizability of CLIG.

\begin{table}[t]
    \centering
    \caption{Online A/B test result for MG-E. Values with asteroids are statistically significant ($p<0.05$) results.}
    \begin{tabular}{c|c|c}
        \toprule
        metrics & MG-E(3 model) vs. baseline & avgImpr \\
        \midrule
        watch time & +0.481\%* & +0.160\%\\
        effective view & +0.416\%* & +0.139\%\\
        like count & +0.220\%* & +0.073\%\\
        share count & +0.375\%* & +0.125\%\\
        follow count & +0.321\%* & +0.107\%\\
        \midrule
        category coverage & +0.012\% & +0.004\%\\
        \bottomrule
    \end{tabular}
    \label{tab: online_result_mg_e}
\end{table}

\begin{table}[t]
    \centering
    \caption{Online A/B test result for CLIG extension. Values with asteroids are statistically significant ($p<0.05$) results.}
    \begin{tabular}{c|c}
        \toprule
        metrics & MG-E + CLIG vs. MG-E \\
        \midrule
        watch time & +0.031\% \\
        effective view & +0.308\%*\\
        like count & +0.009\% \\
        share count & +0.481\%* \\
        follow count & +0.309\% \\
        \midrule
        item coverage & +11.538\%*\\
        intra-list diversity & +11.248\%*\\
        category coverage & +0.255\%*\\
        \bottomrule
    \end{tabular}
    \label{tab: online_result_clig}
\end{table}

\subsubsection{Generator Selection Bias}

\begin{figure}[t]
    \centering
    \includegraphics[width=0.9\linewidth]{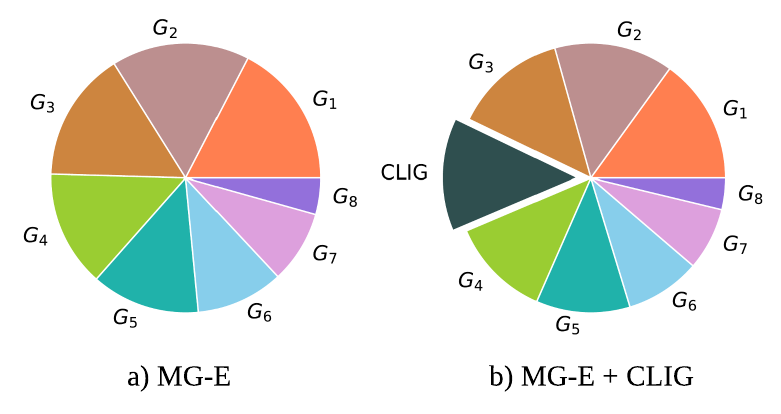}
    \caption{Change of list selection bias when adding CLIG. 
    }
    \label{fig: selection_bias}
\end{figure}

To further verify the effectiveness of the CLIG extension, we observe the evaluator's selection frequency of the generators and present the comparison result in Figure \ref{fig: selection_bias}, corresponding to the comparison in Table \ref{tab: online_result_clig}.
We can see that CLIG can rank in the 4th position of all the 9 generators taking relatively 18\% of the list exposure, which indicates that it is accurate and sufficiently different from others.
We also observe that other generators' exposure scales down at relatively the same rate, indicating an evenly distributed affection, which also validates the CLIG's intuition of being different from other generators.
In addition to the ever-changing model parameters, we remind readers that this generator selection bias also affects the online learning data stream, while adding a new generator would take away other generators' chances to contribute to the data stream for certain types of requests.
To circumvent this selection bias, we adopt an $\epsilon$-greedy strategy that allows the evaluator to randomly select a candidate list as the recommendation with a small probability $\epsilon$.

\section{Conclusion}

In this work, we illustrate that the multi-generator-evaluator framework takes advantage of the ensemble effect and consistently outperforms state-of-the-art single-generator models (with or without evaluator).
Additionally, we verify that the core list-wise diversity measure to the ensemble effect can be achieved by the list comprehensiveness metric, and derive an automatic generator finding method, CLIG.
The CLIG technique is generalizable to different backbone generators and can iteratively increase the number of generators to further improve the performance.
In practice, one should balance the computational cost of the included generators and the decreasing improvement of the recommendation performance.
Yet, coming with the autonomous generator finding strategy is the unknown inference logic behind each reranking decision, which may require further investigation, especially in scenarios where the reasons for recommendations are important.

\section{Appendix}



\appendix

\section{Auto-regressive Kernel}\label{ap: autoregressive_kernel}

\begin{figure}[t]
    \centering
    \includegraphics[width=\linewidth]{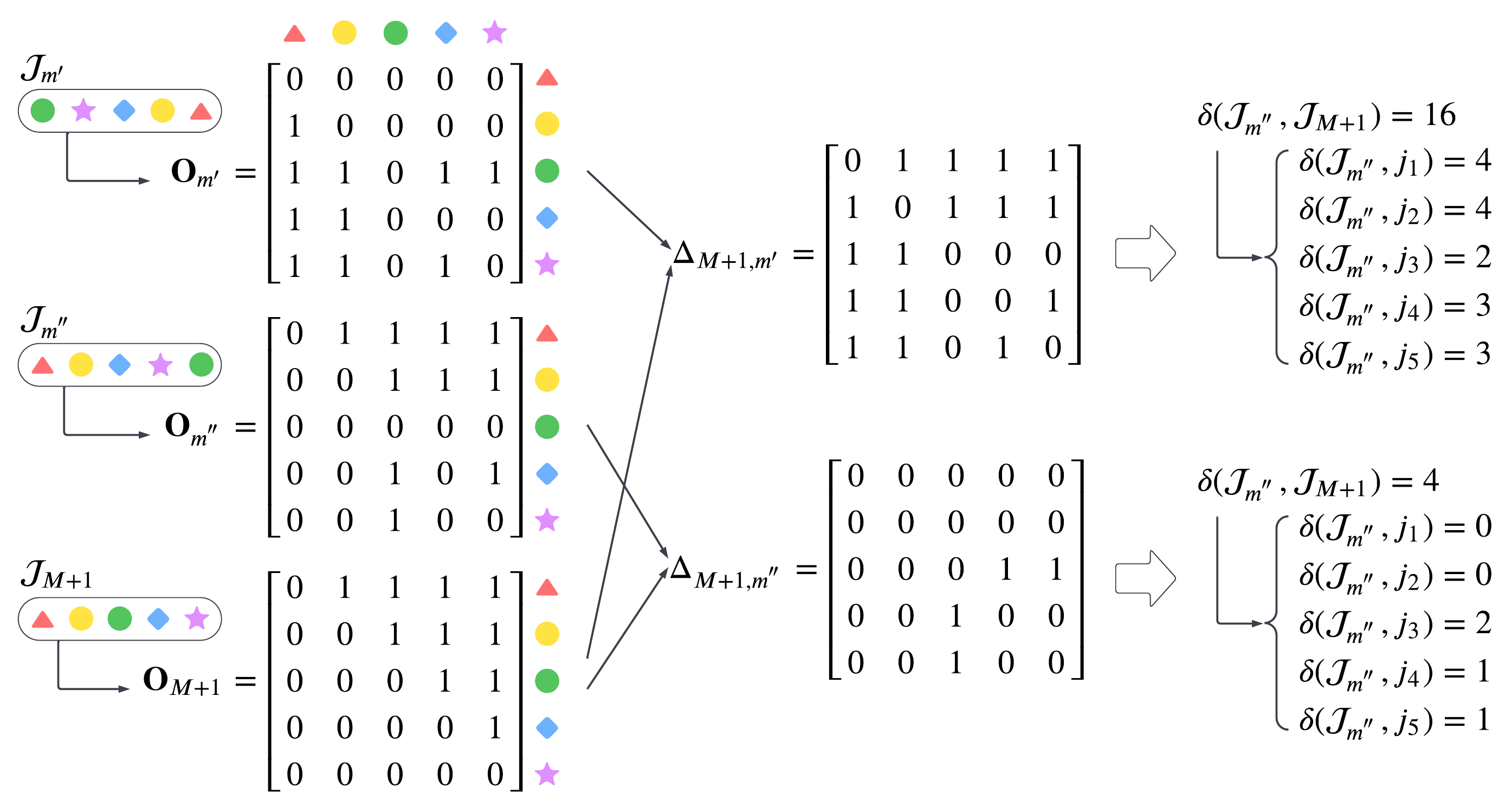}
    \caption{Example of partial order matrix computation of POD kernel. The new $M+1$-th list is compared with two other lists and the list is closer to $\mathcal{J}_{m^{''}}$ compared to $\mathcal{J}_{m^\prime}$.}
    \label{fig: comp_ar_kernel_matrix_example}
\end{figure}

When calculating the GC loss Eq.\eqref{eq: comp_ar} for autoregressive models, we suggest using the item-wise metric $\delta(\mathcal{J}_{m^\prime},j_t)$ rather than the standard list-wise GC metric $\delta(\mathcal{J}_{m^\prime}, \mathcal{J}_{M+1})$ as the guidance to each $P_\theta(j_t|\cdot)$, since items may contribute different magnitudes of ranking diversities.
For the EBD kernel, we can simply set $\delta(\mathcal{J}_{m^\prime},j_t) = 0.5(1-cos(E(\mathcal{J}_{m^\prime}), E(j_t)))$.
For the POD kernel, we can adopt a similar idea and assign the conflicting partial orders related to $j_t$ as the item-wise guidance:
\begin{equation}
\begin{aligned}
    & \delta(\mathcal{J}_{m^\prime},j_t) \\
    = & \sum_{i\in\mathcal{I}}\mathbb{I}\Big[\Big(r(i,\mathcal{J}_{m^\prime})-r(j_t,\mathcal{J}_{m^\prime})\Big)\Big(r(i,\mathcal{J}_{M+1})-r(j_t,\mathcal{J}_{M+1})\Big)<0\Big]
\end{aligned}
\end{equation}
With the GPU-based parallel computing available, we can efficiently calculate this metric through the partial order matrix formulation.
Specifically, we first construct the partial order matrix $\mathbf{O}_m\in\{0,1\}^{n\times n}$ for each list $\mathcal{J}_m$, where $\mathbf{O}_m[i][j] = \mathbb{I}[r(i,\mathcal{J}_m) < r(j,\mathcal{J}_m)]$ represents the order between $i$ and $j$.
One way to efficiently obtain this matrix for a given list is finding the ranks in $\mathcal{J}_m$ for each item in $\mathcal{I}$ as a row vector $\textbf{r}_m = [r(i_1,\mathcal{J}_m),\cdots$ $,r(i_n,\mathcal{J}_m)]$, then cross-subtract with itself and check if each cell is smaller than zero (i.e., $\mathbf{O}_m = \mathbb{I}[(\mathbf{r}_m^\top - \mathbf{r}_m)<0]$).
Then, we define the matrix subtraction $\Delta_{m^\prime, m}=\mathbf{O}_{m^\prime} - \mathbf{O}_{m}$ and calculate the list-wise kernel for non-autoregressive models (in Eq.\eqref{eq: comp}) as:
\begin{equation}
    \delta(\mathcal{J}_{m^\prime}, \mathcal{J}_{M+1}) = \|\Delta_{M+1,m^\prime}\|_1
\end{equation}
and the item-wise kernel for autoregressive models (in Eq.\eqref{eq: comp_ar}) as:
\begin{equation}
    \delta(\mathcal{J}_{m^\prime}, j_t) = \|\Delta_{M+1,m^\prime}[j_t][:]\|_1
\end{equation}
We present the details of the process through a detailed example in Figure \ref{fig: comp_ar_kernel_matrix_example}.
Additionally, the item-wise POD kernel can be considered as the decomposition of the list-wise kernel, i.e., $\|\Delta_{M+1,m^\prime}\|_1 = \sum_{j_t}\|\Delta_{M+1,m^\prime}[j_t][:]\|_1$, which provides a better alignment between the loss of autoregressive and non-autoregressive models.
In contrast, the nonlinear cosine operation does not possess this property, so one may observe significantly different behaviors between Eq.\eqref{eq: comp} and Eq.\eqref{eq: comp_ar} using the EBD kernel.

\newpage

\bibliographystyle{ACM-Reference-Format}
\balance
\bibliography{main}

\end{document}